\documentclass[runningheads]{llncs}
\usepackage{graphicx}
\usepackage{multirow}
\usepackage{color,soul}
\usepackage{amsmath}
\usepackage{amssymb}
\usepackage{cite}
\usepackage{todonotes}
\usepackage{hyperref}
\begin{document}
\title{Controllable cardiac synthesis via disentangled anatomy arithmetic}

%
\author{Spyridon Thermos\inst{1}\index{Thermos, Spyridon} \and
Xiao Liu\inst{1}\index{Liu, Xiao} \and
Alison O'Neil\inst{1,3}\index{O'Neil, Alison} \and
Sotirios A. Tsaftaris\inst{1,2}\index{Tsaftaris, Sotirios A.} }

\authorrunning{S. Thermos et al.}
%
\institute{School of Engineering, University of Edinburgh, Edinburgh EH9 3FB, UK \and
The Alan Turing Institute, London NW1 2DB, UK \and
Canon Medical Research Europe, Edinburgh EH6 5NP, UK\\
\email{\{SThermos,Xiao.Liu,S.Tsaftaris\}@ed.ac.uk} \quad
\email{Alison.ONeil@mre.medical.canon}}
%

\maketitle              
\begin{abstract}
Acquiring annotated data at scale with rare diseases or conditions remains a challenge. 
It would be extremely useful to have a method that controllably synthesizes images that can correct such underrepresentation. Assuming a proper latent representation, the idea of a ``latent vector arithmetic'' could offer the means of achieving such synthesis. 
A proper representation must encode the fidelity of the input data, preserve invariance and equivariance, and permit arithmetic operations. Motivated by the ability to disentangle images into \emph{spatial} anatomy (tensor) factors and accompanying imaging (vector) representations, we propose  a  framework  termed ``disentangled anatomy arithmetic", in which a generative model learns to combine anatomical factors of different input images such that when they are re-entangled with the desired imaging modality (\textit{e.g.}~MRI), plausible new cardiac images are created with the target characteristics. To encourage a realistic combination of anatomy factors after the arithmetic step, we propose a localized noise injection network that precedes the generator. Our model is used to generate realistic images, pathology labels, and segmentation masks that are used to augment the existing datasets and subsequently improve post-hoc classification and segmentation tasks. Code is publicly available at \url{https://github.com/vios-s/DAA-GAN}. 

\keywords{Disentangled anatomy arithmetic \and  semantic image synthesis \and cardiac data augmentation.}
\end{abstract}
%
%
%
%
%
\section{Introduction}
Whilst large scale public datasets are available for traditional vision tasks, medical data are difficult to acquire. Even in a large-scale medical training dataset, examples of rare diseases and anatomies are scarce. As a result, generalisation to observations that are not seen during training will be reduced. To increase the diversity of training data and for instance, to increase the incidence of rare characteristics, we would like the ability to mix and match factors that encode these variations~\cite{bengio2013pami} in a controllable way \textit{i.e.}~perform \textit{controllable image synthesis}.

\begin{figure}[t]
    \centering
    \includegraphics[width=0.95\textwidth]{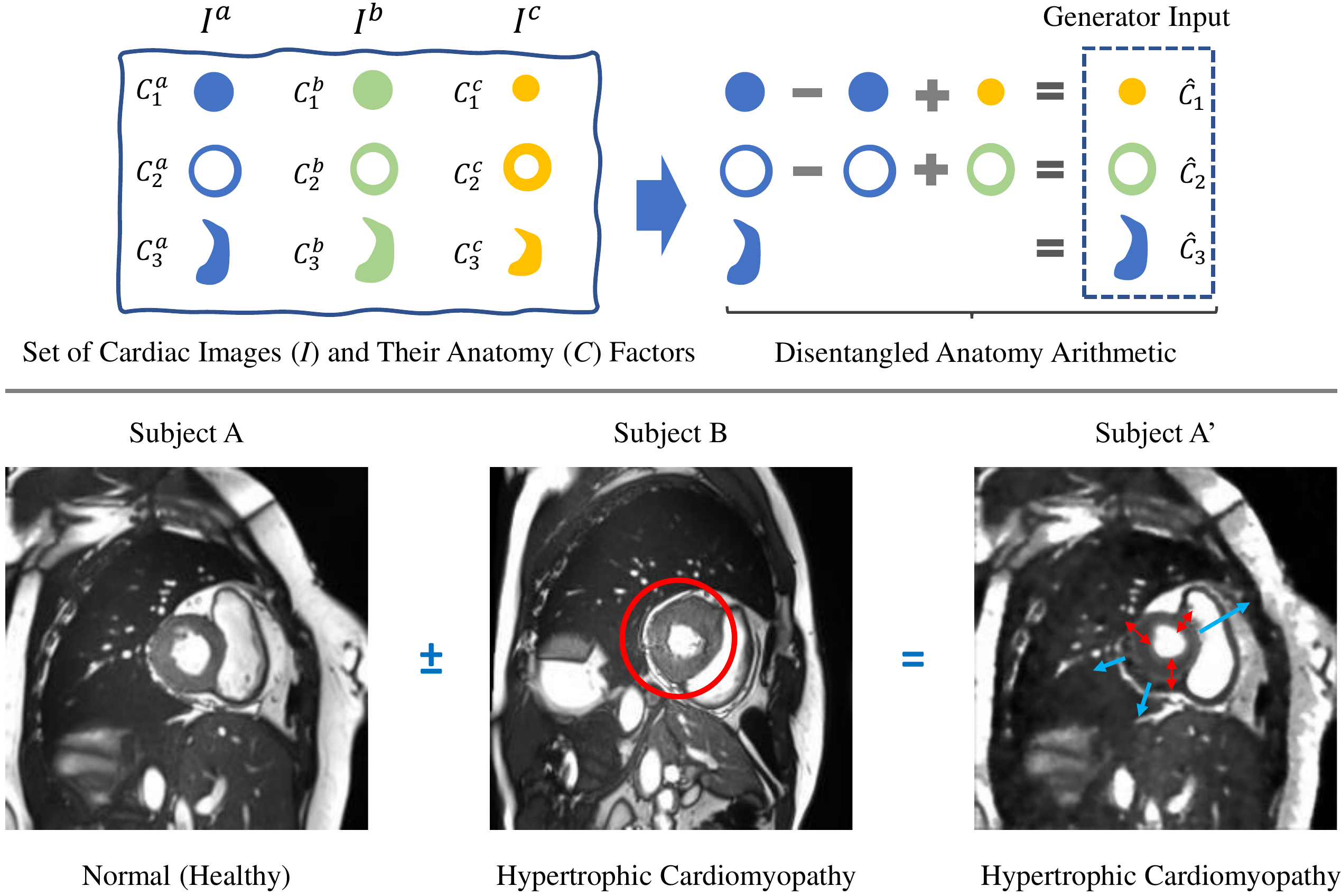}
    \caption{
    \textit{Top}: overview of the ``disentangled anatomy arithmetic" concept illustrated with 3 factors that represent 3 different anatomical parts of the heart (\textit{e.g.}~left/right ventricle and myocardium). \textit{Bottom}: DAA-GAN generated example. Given a healthy Subject A, we aim to generate an image A' which exhibits hypertrophic cardiomyopathy (HCM). We select a Subject B with HCM and remove the anatomical factors from A (i.e. the ones that encode the myocardium and left ventricular cavity) and add the corresponding factors of B (inner part of the red circle). Arrows in A' point to local deformations showing the non-linear abilities of our arithmetic. Arithmetic operations are denoted with \color{blue}$\pm$\color{black}.}
    \label{fig:promo}
\end{figure}

The idea of generating realistic images to augment existing limited data is not new in medical image analysis. Generative Adversarial Networks (GANs)~\cite{goodfellow2014neurips} have been used to generate variants for a given input image based on sampling from a random noise vector. In fact, more recent GAN architectures pursue controllability by disentangling existing factors of variation, conditioning the generation process using semantic priors (\text{e.g.}~segmentation masks)~\cite{guibas2017neuripsw,costa2018tmi,shin2018sashimi,dakai2018miccai,mok2018miccaiw,havaei2020arxiv,li2020sensors} and class labels~\cite{hu2018neuripsw,FridAdar2018neuro}, or learning cross-modality translations~\cite{nie2017miccai,ben2017sashimi,li2019sashimi,chartsias2019stacom,dar2019tmi}. An alternative to relying on sampling from a noise vector for new medical images, is the idea of mixing existing information from different populations (\textit{e.g.}~patients with different anatomical characteristics) to learn the intermediate latent space and generate more realistic data in a more controllable way. A concept that approximates this idea is ``vector arithmetic"~\cite{radford2016iclr}, where existing vector-based latent representations are combined using simple arithmetic operations to produce new images. However, vector representations do not exploit the spatial equivariance of the image content and the respective task (\textit{e.g.}~segmentation) and have shown poor reconstruction quality. An alternative is to use to disentangled representations that use both spatial (tensor) and vector representations to capture factors of variation and permit decomposition of the input in spatially equivariant (and closely to be semantic) and imaging information~\cite{chartsias2019mia,chen2019miccai,yang2019miccai}. 

Herein using disentangled representations we propose the concept of ``disentangled anatomy arithmetic" (DAA), visualized in Fig.~\ref{fig:promo}, that enables controllable image synthesis of plausible images with a \textit{target pathology}, which we show to  be useful for augmenting existing medical training data. We design the DAA-GAN model that learns to combine and transform \emph{spatial} anatomical factors --we provide a visual example of anatomy factors in Fig.~1 of the supplemental-- from input images captured by different vendors or from different populations, and then re-entangle them with the chosen imaging factors (\textit{e.g.}~MRI) to generate unseen intermediate representations. Inspired by recent findings regarding the advantages of introducing spatial stochasticity in the generation process~\cite{karras2019cvpr,alharbi2019cvpr,gabbay2020iclr,alharbi2020cvpr}, we propose a convolutional module for the combined anatomical factor representation transformation, in which structured noise is injected at the spatial locations where the arithmetic operations take place.

Our \textbf{contributions} are to:
\begin{itemize}
    \item Introduce the concept of ``disentangled anatomy arithmetic" on spatial representations of the anatomy.
    \item Propose DAA-GAN, a generative model that to the best of our knowledge, is the first to condition image generation using spatial anatomy factors as semantic priors.
    \item Propose the noise injection module that encourages local deformations to realistically blend the new factors after the arithmetic step.
    \item Evaluate the impact of using DAA-GAN for cardiac data augmentation in the context of a classification and a semantic segmentation post-hoc task.
\end{itemize}

%
%
%
%
%
\section{Generative Model Architecture}
\label{method}
Our model assumes disentangled anatomy and imaging representations as inputs. To obtain them, we use SDNet~\cite{chartsias2019mia} as this model provides binary spatial factors that correspond to the whole anatomy and can be used as semantic priors.

\begin{figure*}[t]
    \centering
    \includegraphics[width=\textwidth]{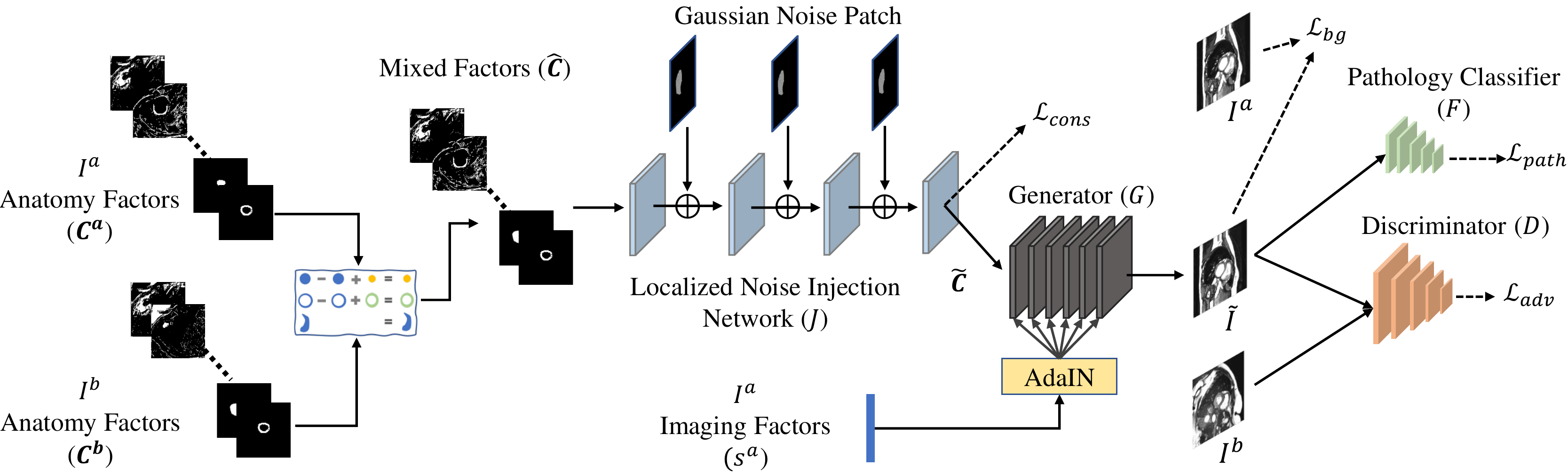}
    \caption{DAA-GAN overview (from left to right): arithmetic operations are performed between anatomical factors $\mathbf{C}^{a}$ and $\mathbf{C}^{b}$ to produce a mixed representation $\mathbf{\hat{C}}$ which is then refined by the noise injection network $\mathcal{J}$. The generator $\mathcal{G}$ receives this refined representation $\mathbf{\tilde{C}}$ and re-entangles it with the the input imaging factors to generate the new image $\tilde{I}$. Finally, a discriminator is responsible to judge if $\tilde{I}$ is real or fake, whilst a pathology classifier assesses if $\tilde{I}$ has the desired cardiac pathology.}
    \label{fig:overview}
\end{figure*}

\textbf{Model overview.} As depicted in Fig.~\ref{fig:overview}, DAA-GAN has 4 distinct steps: 1) We combine anatomy factors, using disentangled anatomy arithmetic, to obtain a new mixed anatomical representation $\mathbf{\hat{C}}$. 2) A noise injection network $\mathcal{J}$ takes $\mathbf{\hat{C}}$ and aims to create a plausible and more refined anatomical representation $\mathbf{\tilde{C}}$. 3) A generator $\mathcal{G}$ reconstructs an image corresponding to $\mathbf{\tilde{C}}$ and a given imaging representation. 4) Two critics, namely a discriminator $\mathcal{D}$ and a pathology classifier $\mathcal{F}$ ensure good image fidelity, but also that the reconstructed image contains the right target characteristics. We proceed detailing these steps.

\textbf{Disentangled anatomy arithmetic.}
As shown in Fig.~\ref{fig:promo} (top), we consider an example with 3 cardiac anatomy populations $\mathbf{C}^{a}, \mathbf{C}^{b}, \mathbf{C}^{c}$ (\textit{e.g.}~patients from 3 imaging populations $a$, $b$, and $c$) with 3 anatomical factors per population, which --when combined-- form a nested structure that corresponds to the heart region. These factors are extracted from $I^{a}, I^{b}, I^{c}$ medical images of dataset $Z_{\sim \{I^{a}, I^{b}, I^{c}\}}$. Based on this setup, we define factor arithmetic between populations any swapping operation between the corresponding factors. Following this example, to create a new $\mathbf{C}^{a}$ anatomy by mixing factors, we first swap $\mathbf{C}_{1}^{a}$ with $\mathbf{C}_{1}^{c}$, and then swap $\mathbf{C}_{2}^{a}$ with $\mathbf{C}_{2}^{b}$. The result is an intermediate $\mathbf{\hat{C}}$ that is used as input to the next module of DAA-GAN. Note that before swapping two factors, assuming upright anatomies, we perform a registration step (warping) to align the swapped-in factor with the center of mass location of the swapped-out one.

\begin{figure}[b]
    \centering
    \includegraphics[width=\textwidth]{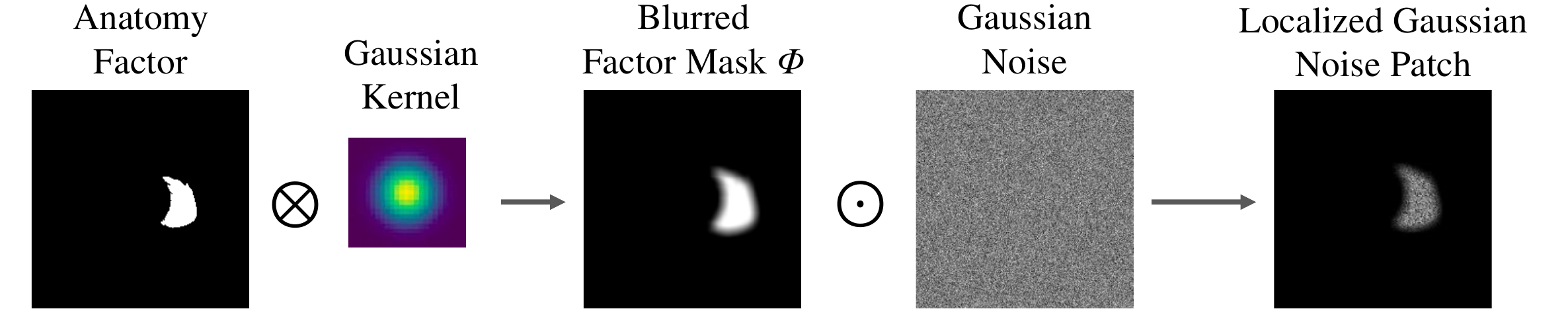}
    \caption{Visualization of the localized Gaussian noise patch generation process. $\otimes$ and $\odot$ denote convolution and element-wise multiplication, respectively.}
    \label{fig:noise}
\end{figure}

\textbf{Noise injection.}
Since cardiac anatomy is a nested structure, the output of the arithmetic step might be non-realistic, \textit{e.g.}~have factors that overlap with each other. This can lead to generated images with ambiguous pathology. We tackle this problem with module $\mathcal{J}$, which receives $\mathbf{\hat{C}}$ and produces a refined representation $\mathbf{\tilde{C}}$.
Inspired by recent work of Karras \textit{et al.}~\cite{karras2019cvpr}, we introduce stochastic variation (as Gaussian noise) at specific spatial locations of each convolutional (CONV) layer's activations. This variation is exploited by the network to cause local deformations around the added (swapped-in) factor(s) in order to preserve the non-overlapping nested structure of the heart (see Fig.~\ref{fig:promo}(bottom)). $\mathcal{J}$ consists of 4 CONV layers, whilst the noise is injected in the form of noise patches (see Fig.~\ref{fig:noise}) added to each CONV layer's activation features in an element-wise fashion. The last CONV layer is followed by a Gumbel-Softmax operator that bounds $\mathbf{\tilde{C}}$ to $[0,1]$.

\textbf{Generator.}
The generator is responsible for the re-entanglement of anatomy and imaging information, and by extension for the generation of a new image $\tilde{I}$. $G$ consists of 4 CONV-ReLU layers followed by a hyperbolic tangent activation function. The first CONV layer receives $\mathbf{\tilde{C}}$ as input, while after each CONV-ReLU block, there is an Adaptive Instance Normalization (AdaIN) layer~\cite{huang2017iccv} that scales and shifts the activations based on the input imaging factors. Since $\mathbf{\tilde{C}}$ has the same dimensions as $I$ and $\tilde{I}$, there is no need for any upsampling.

\textbf{Critics.}
Since for $\tilde{I}$ we have no ground truth we use a discriminator to guide reconstruction. We adopt the architecture of the LSGAN discriminator~\cite{mao2017iccv} for faster convergence, better stability and ultimately better image quality (compared to a variant with spectral normalization layers adopted from ~\cite{miyato2018iclr}). 
Additionally, since we want $\tilde{I}$ to have desired properties, we use a VGG16 model~\cite{simonyan2015vgg} $\mathcal{F}$ to classify the pathology represented in $\tilde{I}$. $\mathcal{F}$ has 7 CONV-BN-ReLU blocks and 3 fully-connected layers followed by a softmax function. Note that $\mathcal{F}$ is pre-trained on the original data, and is then used as a pathology predictor during training of the generative model.

Having presented each module in detail, we now proceed to describe the 4 losses used to train our model as a whole:

\textbf{Adversarial loss ($\mathcal{L}_{adv})$.}
We use an adversarial loss to encourage our model to generate realistic images. We choose to minimize the LSGAN loss~\cite{mao2017iccv} as it is more stable during training and leads to higher quality image generation compared to the traditional GAN loss. Our adversarial loss is defined as:
\begin{equation}
\begin{gathered}
    \mathcal{L}_{adv} = \mathcal{L}_{\mathcal{D}} + \mathcal{L}_{\mathcal{G}},
    \\ 
    \mathcal{L}_{\mathcal{D}} = \frac{1}{2}\mathbb{E}_{I'\sim p(Z)}[(\mathcal{D}(M\cdot I') - 1)^2]\  +
    \frac{1}{2}\mathbb{E}_{\mathbf{C}\sim p(\mathbf{C}_{Z})}[(\mathcal{D}(M\cdot \mathcal{G}(\mathbf{\hat{C}})))^2],
    \\
    \mathcal{L}_{\mathcal{G}} = \frac{1}{2}\mathbb{E}_{\mathbf{C}\sim p(\mathbf{C}_{Z})}[(\mathcal{D}(\mathcal{G}(\mathbf{\hat{C}})) - 1)^2],
\end{gathered}
\end{equation}
where $I'$ is the image that contributes only the anatomy factor(s) $\mathbf{C'}$ which are used to form $\mathbf{\hat{C}}$ (\textit{e.g.}~$I^{b}$ in Fig.~\ref{fig:overview}). $M$ is a binary mask produced by the union of the anatomical factors that contain information about the heart. Note that $M(j)=0$ for the remaining (\textit{i.e.}~non-heart) pixels.

\textbf{Pathology classification ($\mathcal{L}_{path}$).}
Since we know that we add the anatomical factor $C_{k}$ to the mixed representation, we expect to be able to recognize the pathology corresponding to $C_{k}$ in the generated image $\tilde{I}$. To achieve this we minimize the cross entropy loss, defined as $\mathcal{L}_{path} = -\sum_{i=1}^{\Omega}y_{i}\log(p(x_{i}))$,
where $y_{i}$ and $p(x_{i})$ are the ground truth and predicted pathology labels, and $\Omega$ is the number of pathology classes.

\textbf{Anatomical and background consistency ($\mathcal{L}_{cons}$ and $\mathcal{L}_{bg}$).}
$\mathcal{L}_{cons}$ encourages the anatomical factors which are not related with the heart to remain unaltered after the arithmetic and noise injection steps. To find which pixels should not be changed, we use the blurred mask produced during the noise patch generation, denoted as $\Phi(j)$, with $\Phi(j)=0$ for each pixel location $j$ that is not part of the anatomy mixing. We define $\mathcal{L}_{cons} = \frac{1}{N} \sum_{j}(1-\Phi(j)) ||\mathbf{\hat{C}}(j) - \mathbf{\tilde{C}}(j)||_{1}$.
$\mathcal{L}_{bg}$ is the same as $\mathcal{L}_{cons}$, but on  images. Namely, $\mathcal{L}_{bg} = \frac{1}{N} \sum_{j}(1-M(j)) ||(I^{a}(j) - \tilde{I(j)}) ||_{1}$ where $M$ is defined previously, and $N$ is a the total number of pixels.

\textbf{Total loss ($\mathcal{L}_{total}$).} These losses discussed above, are combined as:
\begin{equation}
    \mathcal{L}_{total} = \mathcal{L}_{adv} +
    \mathcal{L}_{path} + \lambda_{1}(\mathcal{L}_{cons} + \mathcal{L}_{bg}),
\end{equation}
where $\lambda_{1}=10$ is a weighting hyperparameter for the consistency losses.

%
%
%
%
%
\section{Experiments}
In this section we present key information on datasets and metrics, and discuss the results of our experiments (see training details in Sec.~1 of the supplemental).

\textbf{Data.}
We use the ACDC~\cite{bernard2018tmi} dataset and data from the M$\&$Ms challenge~\cite{mnms}. \textbf{ACDC} consists of cardiac images acquired from MRI scanners across 100 subjects, and provides pathology annotations (5 classes) for all images and pixel-level segmentation masks for end-systole (ES) and end-diastole (ED) per subject. \textbf{M$\&$Ms} has cardiac-MR images acquired from 4 different (known) sites (domains), across 345 subjects. It provides ED/ES pixel-level segmentation masks annotations for 320 (of 345) subjects and pathology annotations (4 classes) for all images (for more dataset details see Sec.~2 of the supplemental).

\textbf{Metrics.} We approach this in several levels. To measure image \textbf{quality} we compute the Fr\'{e}chet Inception Distance (FID)~\cite{heusel2017neurips}, which quantifies the distance between feature vectors from real and generated images. To quantify the \textbf{utility} of the generated images we study if they help improving the performance of two post-hoc tasks: a) pathology classification, and b) semantic segmentation. For the former, we measure the classification accuracy achieved by a VGG16 model, whilst for the latter, we measure the Dice score~\cite{dice1945measures, sorensen1948method} achieved by a U-Net~\cite{unet} model. The hardest is to assess \textbf{controllability}, \textit{i.e.}~the ability to generate images that faithfully create combinations of anatomical factors. We approximate this by examining the existence of the added or removed pathologies. We train two VGG16 classifiers, one 5-class for ACDC data and one 4-class for M$\&$Ms data, and measure the pathology classification accuracy on the generated images.  

\begin{table}[t]
\begin{center}
\footnotesize
\begin{tabular}{| l | c c | c c | c c | c c |}
    \hline
    & \multicolumn{2}{c|}{\textbf{Balanced}} & \multicolumn{2}{c|}{\textbf{Un/ed Class}} & \multicolumn{2}{c|}{\textbf{Un/ed Vendor}} & \multicolumn{2}{c|}{\textbf{Image Qual.}} \\
    \hline
    \multirow{2}{*}{\textbf{Generation}} & \multicolumn{2}{c|}{\textbf{ACDC}} & \multicolumn{2}{c|}{\textbf{M$\&$Ms}} & \multicolumn{2}{c|}{\textbf{M$\&$Ms}} & \multirow{2}{*}{\textbf{ACDC}} & \multirow{2}{*}{\textbf{M$\&$Ms}}\\
    & \textbf{Acc.}& \textbf{Dice}& \textbf{Acc.} & \textbf{Dice} & \textbf{Acc.} & \textbf{Dice} & & \\
    \hline
    \emph{Original data} & 88.3$_{1.6}$ & 84.9$_{5.0}$ & 82.7$_{0.6}$ & 83.5$_{3.7}$ & 82.7$_{0.6}$ & 83.6$_{3.6}$ & n/a & n/a\\
    \hline
    Ours   & \textbf{91.4}$^{*}_{1.4}$ & \textbf{86.5}$^{**}_{4.7}$ & \textbf{86.0}$^{*}_{0.8}$ & \textbf{85.2}$^{**}_{4.1}$ & \textbf{85.8}$^{*}_{0.7}$ & \textbf{84.7}$^{**}_{4.0}$ &  17.2 & 24.8 \\ 
    \  /without {$\mathcal{F}$}  & 89.0$_{1.6}$ & 85.3$_{5.5}$ & 84.1$_{0.8}$ & 83.9$_{4.0}$ & 84.2$_{0.5}$ & 83.9$_{3.7}$ & 18.8 & 26.3 \\ 
    \  /without {$\mathcal{J}$}  & 89.3$_{1.5}$ & n/a & 84.3$_{0.6}$ & n/a & 84.5$_{0.7}$ & n/a & 21.5 & 27.8 \\ 
    \  /without {$\mathcal{F}$}, $\mathcal{J}$ & 88.7$_{1.6}$ & n/a & 83.1$_{0.8}$ & n/a & 82.9$_{0.7}$ & n/a & 22.5 & 31.2 \\
    \hline
    AC-GAN~\cite{odena2017icml} & 89.2$_{1.2}$ & n/a & 83.9$_{0.7}$ & n/a & 83.0$_{0.8}$ & n/a & \textbf{17.1} & \textbf{24.7} \\
    SPADE~\cite{park2019cvpr} & 88.3$_{1.5}$ & n/a & 83.5$_{0.5}$ & n/a & 83.3$_{0.7}$ & n/a & 20.2 & 28.7 \\ 
\hline
\end{tabular}
\end{center}
\caption{Comparing data augmentation methods in the context of the 4 questions defined in the Results section (see text for details).
We report average (standard deviation as subscript) classification accuracy (Acc.) and segmentation performance as average Dice score. ``n/a" denotes not applicable experiment. * and ** denote significant improvement over the 2$^{nd}$ best method with $p<0.05$ and $p<0.1$ (Wilcoxon non-parametric test), respectively.} 
\label{tab:cardiac}
\end{table}

\textbf{Setup.}
We generate data with our DAA-GAN; SPADE~\cite{park2019cvpr} a model that conditions synthesis using segmentation masks; and AC-GAN~\cite{odena2017icml} a model that conditions synthesis on class rather than semantic priors. Then, we train the two posthoc task-specific models (\textit{i.e.}~VGG16 and U-Net) with the following training sets (applies to both datasets): i) original data (OD), ii) OD augmented with data from DAA-GAN, iii) OD augmented with data from SPADE, and iv) OD augmented with data from AC-GAN. For (i)-(iv) we further augment the training set using traditional intensity (blur, gamma correction) and geometric (crop, rotate, flip, elastic transformation) augmentations for fairness (validation and test sets remain unaltered). Note that for each generated image, we extract and binarize \textit{only} the heart-related $\mathbf{\tilde{C}}$ factors (output of $\mathcal{J}$) and use them as near-ground truth masks for retraining U-Net in the context of post-hoc segmentation.

\textbf{Results.}
To demonstrate the effectiveness of our method, we answer several key questions referring to quantitative results presented in Table~\ref{tab:cardiac}.

\textit{Does DAA-GAN augmentation improve learning on a balanced dataset?} For this experiment we use ACDC, which is balanced in terms of subjects per pathology class. We split the OD into 70\%, 15\%, 15\% subjects for training, validation, and testing, and use DAA-GAN to generate 50 new images for the 5 pathology classes (this corresponds to a 25\% samples increase). These images are picked based on $\mathcal{F}$ confidence. As Table~\ref{tab:cardiac} shows (column ``Balanced'') data generated by our model lead to a 3.1\% absolute classification accuracy improvement compared to using the OD with traditional augmentations, whilst outperforming both AC-GAN and SPADE. Regarding segmentation, compared to only using the OD, our model improves the Dice score by an absolute 1.6\%. Since AC-GAN does not condition on masks and SPADE does not have a mechanism to refine the fused segmentation masks, they cannot be used in this experiment.

\textit{Does DAA-GAN augmentation improve learning of underrepresented classes?} The M$\&$Ms dataset is imbalanced in terms of the Abnormal Right Ventricle (ARV) pathology class, which represents only 5\% of the dataset. We showcase here the impact of our generative model in augmenting underrepresented classes. 
We split the data in a 70-15-15 percentage fashion (as with ACDC), and use DAA-GAN to generate 168 --high $\mathcal{F}$ confidence-- new images (matching the highest represented class) with the ARV pathology by mixing the factor that corresponds to the RV with healthy anatomies from other subjects. We use this new balanced training set that is used to re-train VGG16 and U-Net. As reported in Table~\ref{tab:cardiac}, column ``Un/ed Class'', augmentations improve both accuracy by 3.3\% and Dice by 1.7\% Dice outperforming AC-GAN and SPADE where applicable.

\textit{Does DAA-GAN augmentation improve learning of underrepresented or new domains?} 
As stated above, M$\&$Ms comprises data captured from 4 different sites (A-D), thus 4 different populations. However, data from site D represent only 14\% in the dataset. Here we showcase the augmentation of site D. Thus, in this experiment we aim to balance the training set of M$\&$Ms by augmenting site D data. We adopt the same split percentage with the previous experiments, and augment the original training set by mixing pathological factors from subjects of vendors A-C with anatomies of vendor D. Results in Table~\ref{tab:cardiac}, column ``Un/ed Vendor'', show that by augmenting site D with 101 generated images of high $\mathcal{F}$ confidence value, we improve the two tasks' performance by 3.1\% in accuracy and 1.2\% Dice, respectively, while outperforming AC-GAN and SPADE.

\begin{figure}[t]
    \centering
    \includegraphics[width=\textwidth]{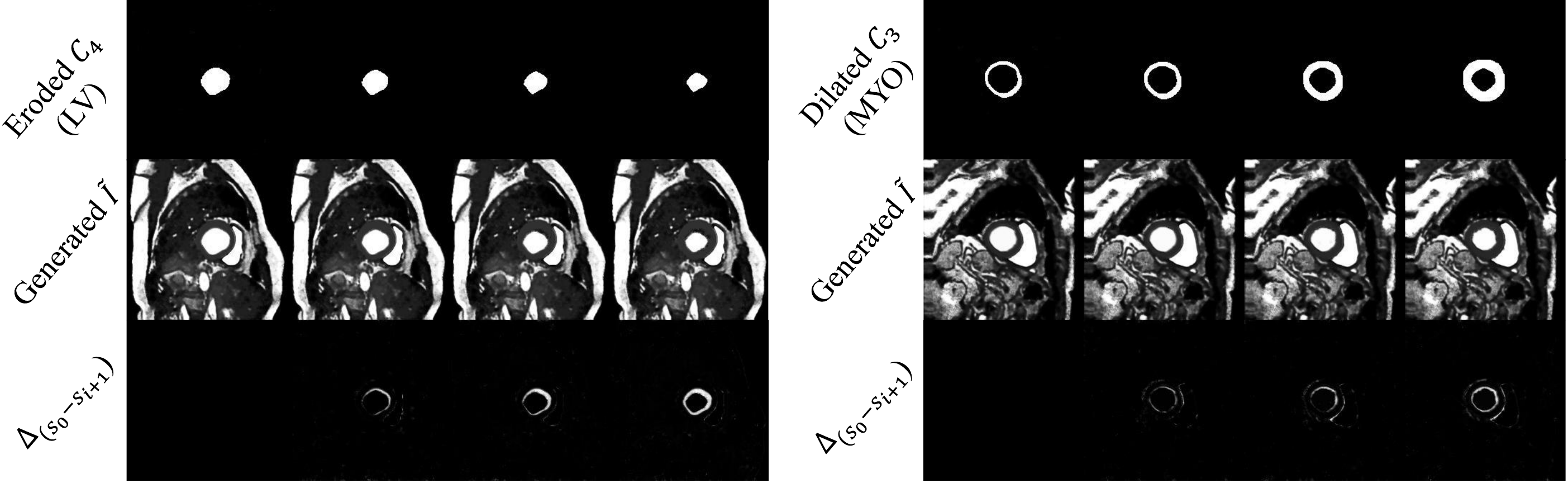}
    \caption{Two examples of anatomy factor traversals. For each step $i$ increases by 3: a) top row depicts the transformed factor $C$, b) middle row depicts the generated images $\tilde{I}$, and c) bottom row shows the factor difference between the current generated image $\tilde{I}$ and the input image. LV, MYO denote the left ventricular cavity and the myocardium.}
    \label{fig:traversals}
\end{figure}

\textit{Does DAA-GAN achieve good image quality?} 
Table~\ref{tab:cardiac} (rightmost) reports generated image quality for each model. Our model outperforms SPADE, which is unsurprising since SPADE has no mechanism to mix semantic priors, thus generating images with unusual heart representations due to overlapping anatomical factors (see examples generated using SPADE in Fig.~3 of the supplemental). AC-GAN, has slightly better FID compared to DAA-GAN in both datasets quality but is the worst model in terms of controllability and utility.

\textit{Can we control DAA-GAN synthesis?} We explore this visually by generating images through manipulation of the anatomical factors of the same subject (\textit{i.e.}~without mixing between subjects). We erode and dilate only a single factor of each subject and generate images based on the altered anatomy. From the examples in Fig.~\ref{fig:traversals}, we observe that the generated cardiac parts do not correspond linearly to the generative factors when approaching extreme cases, \textit{i.e.}~for large kernel size. Thus, we argue that our model's controllability is constrained only by the dataset bias, \textit{i.e.}~``unseen" extreme (possibly unrealistic) combinations. 

\textbf{Ablations.} To evaluate the impact of $\mathcal{J}$ and $\mathcal{F}$ on the augmentation results, we replicate each experiment removing one or both modules. From the results in Table~\ref{tab:cardiac}, we conclude that the two modules have similar contribution to post-hoc classification improvement, $\mathcal{F}$ slightly improves segmentation, whilst we also observe that $\mathcal{J}$ plays an important role in the quality of the generated images.
%
%
%
%
\section{Conclusions}
In this paper we introduced a novel framework for controllable cardiac image synthesis. In particular, we presented a generative model that learns to produce unseen images from existing images using a user-selected combination of spatial anatomy factors. We conducted experiments demonstrating the controllability of the generation process, whilst showcasing the potential of augmenting existing medical data with images generated using the concept of ``disentangled anatomy arithmetic". Future work will focus on extending the capability of our model beyond simple mixing and morphology to richer anatomy arithmetic operations.

\section{Acknowledgement}
This work was supported by the University of Edinburgh, the Royal Academy of Engineering and Canon Medical Research Europe. This work was partially supported by the Alan Turing Institute under the EPSRC grant EP/N510129/1. We thank Nvidia for donating a Titan-X GPU.  S.~A. Tsaftaris acknowledges the support of Canon Medical and the Royal Academy of Engineering and the Research Chairs and Senior Research Fellowships scheme (grant RCSRF1819\textbackslash8\textbackslash25).
%
%
%
%

\bibliographystyle{splncs04}
\bibliography{egbib}

\begin{thebibliography}{10}
\providecommand{\url}[1]{\texttt{#1}}
\providecommand{\urlprefix}{URL }
\providecommand{\doi}[1]{https://doi.org/#1}

\bibitem{alharbi2019cvpr}
Alharbi, Y., Smith, N., Wonka, P.: Latent filter scaling for multimodal
  unsupervised image-to-image translation. In: Proc. CVPR. pp. 1458--1466
  (2019)

\bibitem{alharbi2020cvpr}
Alharbi, Y., Wonka, P.: Disentangled image generation through structured noise
  injection. In: Proc. CVPR. pp. 5134--5142 (2020)

\bibitem{ben2017sashimi}
Ben-Cohen, A., Klang, E., Raskin, S.P., Amitai, M.M., Greenspan, H.: Virtual
  pet images from ct data using deep convolutional networks: Initial results.
  In: Tsaftaris, S.A., Gooya, A., Frangi, A.F., Prince, J.L. (eds.) Proc.
  SASHIMI. pp. 49--57 (2017)

\bibitem{bengio2013pami}
Bengio, Y., Courville, A., Vincent, P.: Representation learning: {A} review and
  new perspectives. IEEE TPAMI  \textbf{35}(8),  1798--1828 (2013)

\bibitem{bernard2018tmi}
Bernard, O., Lalande, A., Zotti, C., {\textit{et al.}}: Deep learning
  techniques for automatic {MRI} cardiac multi-structures segmentation and
  diagnosis: is the problem solved? IEEE TMI  \textbf{37}(11),  2514--2525
  (2018)

\bibitem{mnms}
Campello, V.M., et~al.: Multi-centre, multi-vendor and multi-disease cardiac
  segmentation: {T}he {M$\&$M}s challenge. IEEE TMI (Under Review)  (2020)

\bibitem{chartsias2019mia}
Chartsias, A., Joyce, T., Papanastasiou, G., Semple, S., Williams, M., Newby,
  D.E., Dharmakumar, R., Tsaftaris, S.A.: Disentangled representation learning
  in cardiac image analysis. MIA  \textbf{58} (2019)

\bibitem{chartsias2019stacom}
Chartsias, A., Papanastasiou, G., Wang, C., Stirrat, C., Semple, S., Newby, D.,
  Dharmakumar, R., Tsaftaris, S.A.: Multimodal cardiac segmentation using
  disentangled representation learning. In: Pop, M., Sermesant, M., Camara, O.,
  Zhuang, X., Li, S., Young, A., Mansi, T., Suinesiaputra, A. (eds.) Proc.
  STACOM. pp. 128--137 (2019)

\bibitem{chen2019miccai}
Chen, C., Dou, Q., Jin, Y., Chen, H., Qin, J., Heng, P.A.: Robust multimodal
  brain tumor segmentation via feature disentanglement and gated fusion. In:
  Shen, D., Liu, T., Peters, T.M., Staib, L.H., Essert, C., Zhou, S., Yap,
  P.T., Khan, A. (eds.) Proc. MICCAI. pp. 447--456 (2019)

\bibitem{costa2018tmi}
{Costa}, P., {Galdran}, A., {Meyer}, M.I., {Niemeijer}, M., {Abràmoff}, M.,
  {Mendonça}, A.M., {Campilho}, A.: End-to-end adversarial retinal image
  synthesis. IEEE TMI  \textbf{37}(3),  781--791 (2018)

\bibitem{dar2019tmi}
{Dar}, S.U., {Yurt}, M., {Karacan}, L., {Erdem}, A., {Erdem}, E., {Çukur}, T.:
  Image synthesis in multi-contrast {MRI} with conditional generative
  adversarial networks. IEEE TMI  \textbf{38}(10),  2375--2388 (2019)

\bibitem{dice1945measures}
Dice, L.R.: Measures of the amount of ecologic association between species.
  Ecology  \textbf{26}(3),  297--302 (1945)

\bibitem{FridAdar2018neuro}
Frid-Adar, M., Diamant, I., Klang, E., Amitai, M., Goldberger, J., Greenspan,
  H.: Gan-based synthetic medical image augmentation for increased cnn
  performance in liver lesion classification. Neurocomputing  \textbf{321},
  321--331 (2018)

\bibitem{gabbay2020iclr}
Gabbay, A., Hoshen, Y.: Demystifying inter-class disentanglement. In: ICLR
  (2020)

\bibitem{glorot2010aistats}
Glorot, X., Bengio, Y.: Understanding the difficulty of training deep
  feedforward neural networks. In: Proc. AISTATS (2010)

\bibitem{goodfellow2014neurips}
Goodfellow, I., Pouget-Abadie, J., Mirza, M., Xu, B., Warde-Farley, D., Ozair,
  S., Courville, A., Bengio, Y.: Generative adversarial nets. In: Proc.
  NeurIPS. pp. 2672--2680 (2014)

\bibitem{guibas2017neuripsw}
Guibas, J.T., Virdi, T.S., Li, P.S.: Synthetic medical images from dual
  generative adversarial networks. Advances in Neural Information Processing
  Systems Workshop  (2017)

\bibitem{havaei2020arxiv}
Havaei, M., Mao, X., Wang, Y., Lao, Q.: Conditional generation of medical
  images via disentangled adversarial inference (2020)

\bibitem{heusel2017neurips}
Heusel, M., Ramsauer, H., Unterthiner, T., Nessler, B., Hochreiter, S.: {GAN}s
  trained by a two time-scale update rule converge to a local {N}ash
  equilibrium. In: Proc. NeurIPS. pp. 6626--6637 (2017)

\bibitem{hu2018neuripsw}
Hu, X., Chung, A.G., Fieguth, P., Khalvati, F., Haider, M.A., Wong, A.:
  Prostate{GAN}: {M}itigating data bias via prostate diffusion imaging
  synthesis with generative adversarial networks. NeurIPS Workshop  (2018)

\bibitem{huang2017iccv}
Huang, X., Belongie, S.: Arbitrary style transfer in real-time with adaptive
  instance normalization. In: Proc. ICCV. pp. 1501--1510 (2017)

\bibitem{dakai2018miccai}
Jin, D., Xu, Z., Tang, Y., Harrison, A.P., Mollura, D.J.: {CT-R}ealistic lung
  nodule simulation from 3{D} conditional generative adversarial networks for
  robust lung segmentation. In: Frangi, A.F., Schnabel, J.A., Davatzikos, C.,
  Alberola-L{\'o}pez, C., Fichtinger, G. (eds.) Proc. MICCAI. pp. 732--740
  (2018)

\bibitem{karras2019cvpr}
Karras, T., Laine, S., Aila, T.: A style-based generator architecture for
  generative adversarial networks. In: Proc. CVPR. pp. 4396--4405 (2019)

\bibitem{li2019sashimi}
Li, K., Yu, L., Wang, S., Heng, P.A.: Unsupervised retina image synthesis via
  disentangled representation learning. In: Burgos, N., Gooya, A., Svoboda, D.
  (eds.) Proc. SASHIMI. pp. 32--41 (2019)

\bibitem{li2020sensors}
Li, Q., Yu, Z., Wang, Y., Zheng, H.: Tumorgan: A multi-modal data augmentation
  framework for brain tumor segmentation. Sensors  \textbf{20}(15) (2020)

\bibitem{mao2017iccv}
Mao, X., Li, Q., Xie, H., Lau, R.Y.K., Wang, Z., Smolley, S.P.: Least squares
  generative adversarial networks. In: Proc. ICCV. pp. 2813--2821 (2017)

\bibitem{miyato2018iclr}
Miyato, T., Kataoka, T., Koyama, M., Yoshida, Y.: Spectral normalization for
  generative adversarial networks. In: ICLR (2018)

\bibitem{mok2018miccaiw}
Mok, T.C., Chung, A.C.: Learning data augmentation for brain tumor segmentation
  with coarse-to-fine generative adversarial networks. In: Proc. MICCAI Brain
  Lesion Workshop. pp. 70--80 (2018)

\bibitem{nie2017miccai}
Nie, D., Trullo, R., Lian, J., Petitjean, C., Ruan, S., Wang, Q., Shen, D.:
  Medical image synthesis with context-aware generative adversarial networks.
  In: Descoteaux, M., Maier-Hein, L., Franz, A., Jannin, P., Collins, D.L.,
  Duchesne, S. (eds.) Proc. MICCAI. pp. 417--425 (2017)

\bibitem{odena2017icml}
Odena, A., Olah, C., Shlens, J.: Conditional image synthesis with auxiliary
  classifier {GAN}s. In: Proc. ICML. pp. 2642--2651 (2017)

\bibitem{park2019cvpr}
Park, T., Liu, M.Y., Wang, T.C., Zhu, J.Y.: Semantic image synthesis with
  spatially-adaptive normalization. In: Proc. CVPR. pp. 2337--2346 (2019)

\bibitem{pytorch}
Paszke, A., Gross, S., Chintala, S., Chanan, G., Yang, E., DeVito, Z., Lin, Z.,
  Desmaison, A., Antiga, L., Lerer, A.: Automatic differentiation in {PyTorch}
  (2017)

\bibitem{radford2016iclr}
Radford, A., Metz, L., Chintala, S.: Unsupervised representation learning with
  deep convolutional generative adversarial networks. In: ICLR (2016)

\bibitem{unet}
Ronneberger, O., Fischer, P., Brox, T.: U-{N}et: {C}onvolutional networks for
  biomedical image segmentation. In: Navab, N., Hornegger, J., Wells, W.M.,
  Frangi, A.F. (eds.) Proc. MICCAI. pp. 234--241 (2015)

\bibitem{shin2018sashimi}
Shin, H.C., Tenenholtz, N.A., Rogers, J.K., Schwarz, C.G., Senjem, M.L.,
  Gunter, J.L., Andriole, K.P., Michalski, M.: Medical image synthesis for data
  augmentation and anonymization using generative adversarial networks. In:
  Gooya, A., Goksel, O., Oguz, I., Burgos, N. (eds.) Proc. SASHIMI. pp. 1--11
  (2018)

\bibitem{simonyan2015vgg}
Simonyan, K., Zisserman, A.: Very deep convolutional networks for large-scale
  image recognition. In: ICLR (2015)

\bibitem{sorensen1948method}
S{\o}rensen, T.: A method of establishing groups of equal amplitude in plant
  sociology based on similarity of species content and its application to
  analyses of the vegetation on danish commons. Royal Danish Academy of
  Sciences and Letters  \textbf{5}(4),  1--34 (1948)

\bibitem{yang2019miccai}
Yang, J., Dvornek, N.C., Zhang, F., Chapiro, J., Lin, M., Duncan, J.S.:
  Unsupervised domain adaptation via disentangled representations:
  {A}pplication to cross-modality liver segmentation. In: Shen, D., Liu, T.,
  Peters, T.M., Staib, L.H., Essert, C., Zhou, S., Yap, P.T., Khan, A. (eds.)
  Proc. MICCAI. pp. 255--263 (2019)

\end{thebibliography}

\appendix
\section{Experimental Setting}
Both $\mathcal{J}$ and $\mathcal{D}$ were trained for 90 epochs, using the Adam optimizer with $\beta_{1}=0$, $\beta_{2}=0.999$, and a learning rate of 0.0001. Our pathology classification model $\mathcal{F}$ was pre-trained separately using the Adam optimizer with $\beta_{1}=0.9$, $\beta_{2}=0.999$, and a learning rate of 0.0001, with classification accuracy in the corresponding validation set as the early stopping criterion. We use the the pre-trained decoder of SDNet as our generator $\mathcal{G}$. During training of the generative model, the weights of $\mathcal{F}$ and $\mathcal{G}$ are frozen\footnote{For the ablation study, when we remove $\mathcal{J}$, we instead fine-tune $\mathcal{G}$ during generative model training, with learning rate set to 0.0001.}. Finally, the weights of $\mathcal{D}$ were initialized using the Xavier method~\cite{glorot2010aistats}. DAA-GAN is implemented using PyTorch~\cite{pytorch}, while all experiments were conducted on an Nvidia GTX 1080 Ti graphics processor. 

For all experiments, when performing anatomy arithmetic we allow only one pathology per subject. That is, we may add a factor exhibiting a certain pathology into an otherwise healthy subject, or swap one factor for another factor of the same pathology, but we do not combine two factors with different pathologies in the same subject. During inference, our model uses $\mathcal{J}$ and $\mathcal{G}$ to generate an image with the targeted pathology and $\mathcal{F}$ to generate the pathology label, while $\mathcal{D}$ is discarded.

\begin{figure}[t]
    \centering
    \includegraphics[width=0.75\columnwidth]{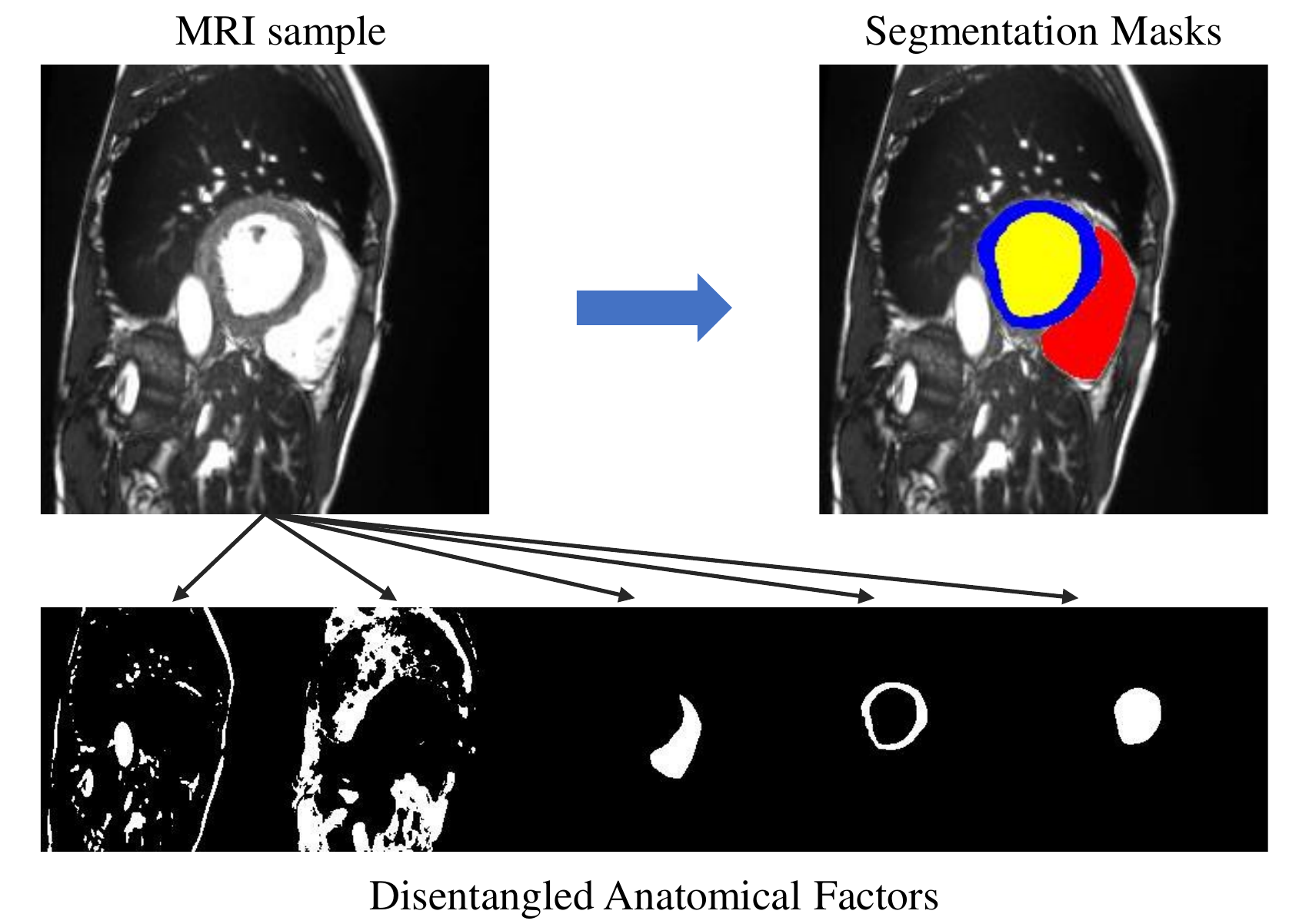}
    \caption{Example of MRI sample from ACDC cardiac dataset (top left), the predicted segmentation masks of the ROI (top right), and the 5 (out of 12) most semantic disentangled anatomical factors. Blue, yellow, and red show the segmentation prediction for the myocardium (MYO), left ventricle (LV) and right ventricle (RV), respectively.}
    \label{fig:intro_mri}
\end{figure}

\section{Dataset Details}
\label{sec:datasets}
\textbf{ACDC} pathology class annotations: a) normal (NOR), b) myocardial infarction (MINF), c) dilated cardiomyopathy (DCM), d) hypertrophic cardiomyopathy (HCM), and e) abnormal right ventricle (ARV). ACDC segmentation annotations (3 semantic classes): left ventricular cavity (LV), myocardium (MYO) of the LV, and right ventricle (RV). All images are resampled to 1.37mm$^2$/pixel resolution and cropped to $224\times 224$ pixels.

\noindent\textbf{M$\&$Ms} (see also \url{https://www.ub.edu/mnms/}) comprises class annotations for 19 pathologies. However, 15 of them represent the $12\%$ of the subjects that corresponds to 1-3 subjects per class, thus we choose to experiment on the 4 most dominant classes that are: a) NOR, b) DCM, c) HCM, and d) ARV. Note that ARV in our experiments is the underrepresented class, but with 26 subjects. M$\&$M segmentation annotations are identical with ACDC ones. All images are resampled to 1.2mm$^2$/pixel resolution and cropped to $224\times 224$ pixels.

For both datasets during both model training and evaluation we combine pairs of MR images from the same heart axis slice level.

\begin{figure}[t]
    \centering
    \includegraphics[width=0.9\columnwidth]{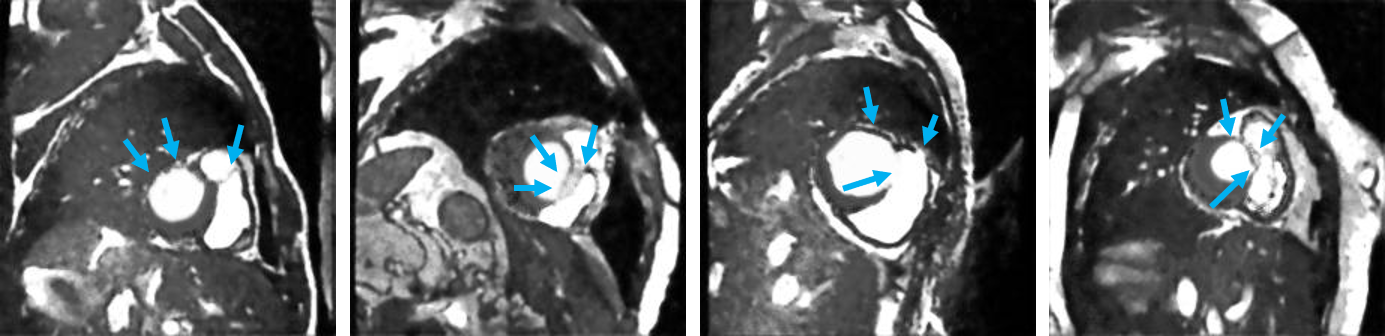}
    \caption{Four SPADE-generated images using overlapped anatomical factors as input. Arrows point to the areas where the performed arithmetic operations lead to such overlaps.}
    \label{fig:spade}
\end{figure}

\end{document}